\documentclass[onecolumn,journal]{IEEEtran}
\usepackage{CJK}
\usepackage{cite}
\usepackage{amsmath,bm}
\usepackage{amsfonts}
\usepackage{amssymb}
\usepackage{amsthm}
\usepackage{multirow,bigstrut,threeparttable}
\newtheorem{theorem}{Theorem}

\usepackage{graphicx}
\usepackage{pdfpages}
\usepackage{bbm}
\usepackage{epstopdf}
\usepackage{graphicx}
\usepackage{subfigure}
\usepackage{amssymb,amsfonts,amsmath}
\usepackage{multirow,bigstrut,threeparttable}
\usepackage{graphicx}
\usepackage{pdfpages}
\newtheorem{question}{Question}
\usepackage{bbm}

\begin{document}

\title{Beyond Maximum Likelihood: from Theory to Practice} %entropy and 
%% Molecular, Murcia, Spain}, \and Franklin Sonnery\affil{2}{}}

\author{Jiantao~Jiao,~\IEEEmembership{Student Member,~IEEE},~Kartik~Venkat,~\IEEEmembership{Student Member,~IEEE},~Yanjun~Han,~\IEEEmembership{Student Member,~IEEE}, and Tsachy~Weissman,~\IEEEmembership{Fellow,~IEEE}% <-this % stops a space
\thanks{Jiantao Jiao, Kartik Venkat, and Tsachy Weissman are with the Department of Electrical Engineering, Stanford University, CA, USA. Email: \{jiantao,kvenkat,tsachy\}@stanford.edu}. % <-this % stops a space
\thanks{Yanjun Han is with the Department of Electronic Engineering, Tsinghua University, Beijing, China. Email: hanyj11@mails.tsinghua.edu.cn}
}

%----------------------------------------------------------------------------------------

\maketitle % The \maketitle command is necessary to build the title page

%----------------------------------------------------------------------------------------
%	ABSTRACT, KEYWORDS AND ABBREVIATIONS
%----------------------------------------------------------------------------------------

\begin{abstract}
Maximum likelihood is the most widely used statistical estimation technique. Recent work by Jiao, Venkat, Han, and Weissman~\cite{Jiao--Venkat--Han--Weissman2014minimax} introduced a general methodology for the construction of estimators for functionals in parametric models, and demonstrated improvements - both in theory and in practice - over the maximum likelihood estimator (MLE), particularly in high dimensional scenarios involving parameter dimension comparable to or larger than the number of samples. This approach to estimation, building on results from approximation theory, is shown to yield minimax rate-optimal estimators for a wide class of functionals, implementable with modest computational requirements. In a nutshell, a message of this recent work is that,  for a wide class of functionals, the performance of these essentially optimal estimators with $n$ samples is comparable to that of the MLE with $n \ln n$ samples.

In the present paper, we highlight the applicability of the aforementioned methodology to statistical problems beyond functional estimation, and show that it can yield substantial gains. For example, we demonstrate that for learning tree-structured graphical models, our approach achieves a significant reduction of the required data size compared with the classical Chow--Liu algorithm, which is an implementation of the MLE, to achieve the same accuracy. The key step in improving the Chow--Liu algorithm is to replace the empirical mutual information with the estimator for mutual information proposed in~\cite{Jiao--Venkat--Han--Weissman2014minimax}. Further, applying the same replacement approach to classical Bayesian network classification, the resulting classifiers uniformly outperform the previous classifiers on 26 widely used datasets.
\end{abstract}

\section{Introduction}
Maximum likelihood emerged in modern form 90 years ago in a series of remarkable papers by Fisher \cite{Fisher1922mathematical, Fisher1925theory, Fisher1934two}, and has since risen to prominence as the most widely used statistical estimation technique. As evidence of its ubiquity, the Google Scholar %\cite{GoogleScholar} 
search query ``Maximum Likelihood Estimation'' yields approximately $2,570,000$ articles, patents and books. Indeed, in his response to Berkson \cite{Berkson1980minimum} in 1980, Efron explains the popularity of maximum likelihood:

\begin{center}
\parbox{.85\textwidth}{~~\emph{  ``The appeal of maximum likelihood stems from its universal applicability, good mathematical properties, by which I refer to the standard asymptotic and exponential family results, and generally good track record as a tool in applied statistics, a record accumulated over fifty years of heavy usage.
"}}
\end{center} 

Over the years, the following folk theorem seems to have been tacitly accepted by applied scientists:

\begin{theorem}[``Folk Theorem'']\label{thm.folk}
For a finite dimensional parametric estimation problem, it is ``good'' to employ the MLE.
\end{theorem}

From the perspective of mathematical statistics, however, maximum likelihood is by no means sacrosanct. As early as in 1930, in his letters to Fisher, Hotelling raised the possibility of the MLE performing poorly \cite{Stigler2007epic}. Subsequently, various examples showing that the performance of the MLE can be significantly improved upon, have been proposed in the literature, cf. Le Cam~\cite{LeCam1990maximum} for an excellent overview. However, as Stigler~\cite[Sec. 12]{Stigler2007epic} discussed in his 2007 survey, while these early examples created a flurry of excitement, for the most part they were not seen as debilitating to the fundamental theory. Perhaps because these examples did not provide a systematic methodology for improving the MLE. 

In 1956, Stein~\cite{Stein1956inadmissibility} observed that in the Gaussian location model $X \sim \mathcal{N}(\theta, I_p)$ (where $I_p$ is the $p\times p$ identity matrix), the MLE for $\theta$, $\hat{\theta}^{\textrm{MLE}} = X$ is inadmissible \cite[Chap. 1]{Lehmann--Casella1998theory} when $p\geq 3$. Later, James and Stein~\cite{James--Stein1961estimation} showed that an estimator that appropriately shrinks the MLE towards zero achieves uniformly lower $L_2$ risk compared to the risk of the MLE. The \emph{shrinkage} idea underlying the James--Stein estimator has proven extremely fruitful for statistical methodology, and has motivated further milestone developments in statistics, such as wavelet shrinkage~\cite{Donoho--Johnstone1994ideal}, and compressed sensing~\cite{Candes--Tao2006near,Donoho2006compressed}. 

One interpretation of the shrinkage idea is that, when one desires to estimate a high dimensional parameter, the MLE may have a relatively small bias compared to the variance. Shrinking the MLE introduces an additional bias, but reduces the overall risk by reducing the variance substantially. A natural question now arises: What about situations wherein the bias is the dominating term? Does there exist an analogous methodology for improving over the performance of the MLE in such scenarios? A precedent to this line of questioning can be found in the 1981 Wald Memorial Lecture by Efron~\cite{Efron1982maximum} entitled ``Maximum Likelihood and Decision Theory'': 

\begin{center}
\parbox{.85\textwidth}{~~\emph{  ``%The fact that it (the MLE) automatically estimates all possible parameters $\gamma(f)$ strongly suggests that 
\ldots the MLE can be non-optimal if the statistician has one specific estimation problem in mind. Arbitrarily bad counterexamples, along the line of estimating $e^{\theta}$ from $X \sim \mathcal{N}(\theta,1)$, are easy to construct. Nevertheless the MLE has a good reputation, acquired over 60 years of heavy use, for producing reasonable point estimates. Useful general improvements on the MLE, such as robust estimation, and Stein estimation, are all the more impressive for their rarity. 
"}}
\end{center}

For the aforementioned example, Efron~\cite{Efron1982maximum} argued that the reason the MLE $e^X$ may not be a good estimate for $e^\theta$, is that it has a large \emph{bias}. In particular, the statistician may prefer the uniform minimum variance unbiased estimator (UMVUE), $e^{X-\frac{1}{2}}$ to estimate $e^\theta$. As will be shown later, the bias is usually the dominating term in estimation of functionals of high-dimensional parameters. 

It is worth recalling at this point that the MLE in functional estimation has a powerful refuge: \emph{asymptotic efficiency}. In other words, for a large family of models, as the number of observed samples grows without bound while the parameter dimension remains fixed, the MLE performs optimally, cf. \cite[Chap. 8]{Vandervaart2000}. However, this guarantee ceases to be valid as soon as the parameter dimension and the sample size are comparably large. How should statisticians inform their usage of the MLE in such cases? Notably, the two general improvements of the MLE, namely robust estimation and shrinkage estimation, are not designed to handle functional estimation problems such as the one presented by Efron. However, as Efron himself observed, the statistician cannot always rely on the UMVUE to save the day, since these are generally very hard to compute, and may not always exist \cite[Remark C, Sec. 7]{Efron1982maximum}. Thus, there is a need to address, both in scope and methodology, the improvement over the MLE for problems where the  bias is the leading term. Such a solution could be considered the dual of the idea of shrinkage, since the trade-off between bias and variance is now reversed, i.e., one might want to sacrifice the variance to reduce the bias.  

\begin{question}
Does there exist a systematic methodology for improving MLE in cases where the bias dominates the risk, such as in functional estimation?
\end{question}

Our goal in this paper is to answer this general question. We prescribe a systematic methodology for improving MLE in functional estimation, and demonstrate the potential of this methodology in solving other statistical problems. The key idea in this approach, introduced in \cite{Jiao--Venkat--Han--Weissman2014minimax}, is to estimate not the functional, but an approximation of the functional where the functional is ``non-smooth'', and to use a bias corrected version of the MLE in regions where the functional is ``smooth''. Additionally, the procedure, which is shown to yield minimax rate-optimal schemes for a large family of functional estimation problems, requires nearly no additional computational overhead -- and is therefore implementable in practice.

These results suggest that Theorem~\ref{thm.folk} is far from true in practice. In particular, we demonstrate that, in settings where the parameter dimension is large, the MLE is generally highly sub-optimal. This makes our methodology valuable in practice, especially in the current era of ``big data'' which necessitates going beyond classical asymptotic analysis and considering finitely many samples in high dimensions. Among other influences, our work is inspired by the recent successes of finite-blocklength analysis in information theory~\cite{Polyanskiy--Poor--Verdu2010channel}, and of compressed sensing in statistics~\cite{Candes--Tao2006near, Donoho2006compressed} that have demonstrated the benefit of carefully analyzing finite and practical sample sizes. 

To demonstrate the efficacy of our methodology, we focus on estimating the entropy and mutual information for discrete distributions, and emphasize that a much wider class of functionals are accommodated by the same approach~\cite{Jiao--Venkat--Han--Weissman2014minimax}. The entropy and mutual information are two fundamental information measures with many applications in compression and communication of data~\cite{Cover--Thomas2006}, in statistical decision theory~\cite{Cesa--Lugosi2006}, and in neuroscience~\cite{Nemenman--Shafee--Bialek2001}, among other disciplines. The entropy, due to Shannon~\cite{Shannon1948}, is the amount of information required to describe an object of a given distribution, while the mutual information naturally characterizes the amount of dependence between two random variables~\cite{Jiao--Courtade--Venkat--Weissman2014Justification}. In particular, as we show in detail below, many of the algorithms in machine learning and statistics either explicitly or implicitly involve estimating entropy and mutual information. In light of this, we expect the schemes for estimating entropy and mutual information stemming from~\cite{Jiao--Venkat--Han--Weissman2014minimax} will lead to new insights and performance boosts in various disciplines. 
%with various important variants such as directed information~\cite{Massey90,Jiao--Permuter--Zhao--Kim--Weissman2013}

Specifically, the contributions of this paper are threefold. (i) We demonstrate that the estimation technique developed in \cite{Jiao--Venkat--Han--Weissman2014minimax} results in the optimal sample complexity and minimax rates for estimation of mutual information. We then showcase the benefit of employing this improved estimator in two widely studied applications. (ii) We demonstrate that, for learning the structure of tree graphical models, replacing the empirical mutual information used in the Chow--Liu algorithm~\cite{Chow--Liu1968} with the improved estimator results in significant performance boosts. Notably, the Chow--Liu algorithm \cite{Chow--Liu1968} implements the MLE for this problem. (iii) We advocate the use of this estimator in a Bayesian network classification problem~\cite{Friedman--Geiger--Goldszmidt1997bayesian}, where replacement of the empirical mutual information estimator by our improved estimator yields improvements in the classification error, uniformly over all the popular datasets. 

Thus, one of the main contributions of this paper is in identifying important problems where estimation of information measures is a key component, and highlighting that there not only exist improvements over the MLE, but that these improvements can yield significant performance boosts in the ``downstream applications'', and moreover that they can be achieved with little to no additional computational cost.  

%The rest of the paper is organized as follows. We discuss our methodology of improving MLE and its applications in various statistical problems in the main results section, and discuss some future work in the concluding remarks. 

Logarithms in this paper are assumed in the natural base. 

%------------------------------------------------

\section{Main Results}

\subsection{General methodology of functional estimation}

Recently, we proposed a general methodology of constructing minimax estimators for functionals~\cite{Jiao--Venkat--Han--Weissman2014minimax}, and showed that the MLE is generally far from minimax optimality~\cite{Jiao--Venkat--Weissman2014MLE}. The methodology can be summarized as follows.

Consider estimating functional $F(\theta)$ of a parameter $\theta \in \Theta \subset \mathbb{R}^p$ for an arbitrary experiment $\{P_\theta: \theta \in \Theta\}$. Suppose we are given an unbiased estimator $\hat{\theta}_n$ for $\theta$, where $n$ is the number of observations. Suppose the functional $F(\theta)$ is continuous everywhere, and differentiable except at $\theta \in \Theta_0$. 
\begin{enumerate}
\item \textbf{Classify Regime}: Compute $\hat{\theta}_n$, and declare that we are operating in the ``smooth'' regime if $\|\hat{\theta}_n - \theta_0\| > \Delta_n, \forall \theta_0 \in \Theta_0$, where $\| \cdot \|$ is some distance function. Otherwise declare we are in the ``non-smooth'' regime;
\item {\bf Estimate}:
\begin{enumerate}
\item If $\hat{\theta}_n$ falls in the ``smooth'' regime, use an estimator similar to $F(\hat{\theta}_n)$ to estimate $F(\theta)$;
\item If $\hat{\theta}_n$ falls in the ``non-smooth'' regime, compute the best approximation of the function $F(\theta)$ near $\theta_0 \in \Theta_0$ using polynomials or trigonometric series up to a specified order $K_n$, and estimate this polynomial (or trigonometric polynomial) instead of $F(\theta)$.
\end{enumerate}
\end{enumerate}

The key idea in the above approach is that in the ``smooth'' regime the plug-in estimators will perform reasonably well even non-asymptotically with some adjustments. On the other hand, the ``non-smooth'' regime requires the construction of an estimator specifically designed for the corresponding parameters. It turns out that the correct approach towards estimation in the ``non-smooth'' regime is to estimate, not the functional itself, but a good approximation of it via the closest (in sup norm, to the original functional) polynomial of a fixed order $K_n$. In most scenarios of interest it is simple and natural to construct unbiased estimators for the integer powers of parameters which present themselves in such a representation. In~\cite{Jiao--Venkat--Han--Weissman2014minimax}, we provided tools to determine $\Delta_n$ and $K_n$ in various statistical experiments. Moreover, it was shown that this methodology achieves the minimax rates for estimating the Shannon entropy $H(P) = \sum_{i = 1}^S -p_i \ln p_i$ as well as the functional $F_\alpha(P) = \sum_{i = 1}^S p_i^\alpha$, for any $\alpha>0$.

It is insightful to compare the aforementioned methodology with shrinkage. The rationale behind shrinkage is to significantly reduce the variance at the expense of slightly increasing the bias. However, it has long been observed in the literature on entropy estimation that the bias dominates the $L_2$ risk \cite{Paninski2003}. In our methodology, one significantly reduces the bias at the expense of slightly increasing the variance via the theory of \emph{approximation}. \footnote{The usage of approximation theory in functional estimation has precedent, dating back to \cite{Ibragimov--Nemirovskii--Khasminskii1987some}. Lepski, Nemirovski, and Spokoiny \cite{Lepski--Nemirovski--Spokoiny1999estimation} utilized trigonometric approximation to estimate the $L_r$ norm of a regression function. Cai and Low \cite{Cai--Low2011} used polynomial approximation to estimate the $\ell_1$ norm of a normal mean. } 
% XXX I did not write too much about the 1987 paper of Nemirovski et al. 

\begin{figure}[h]
\centerline{\includegraphics[width=0.8\linewidth]{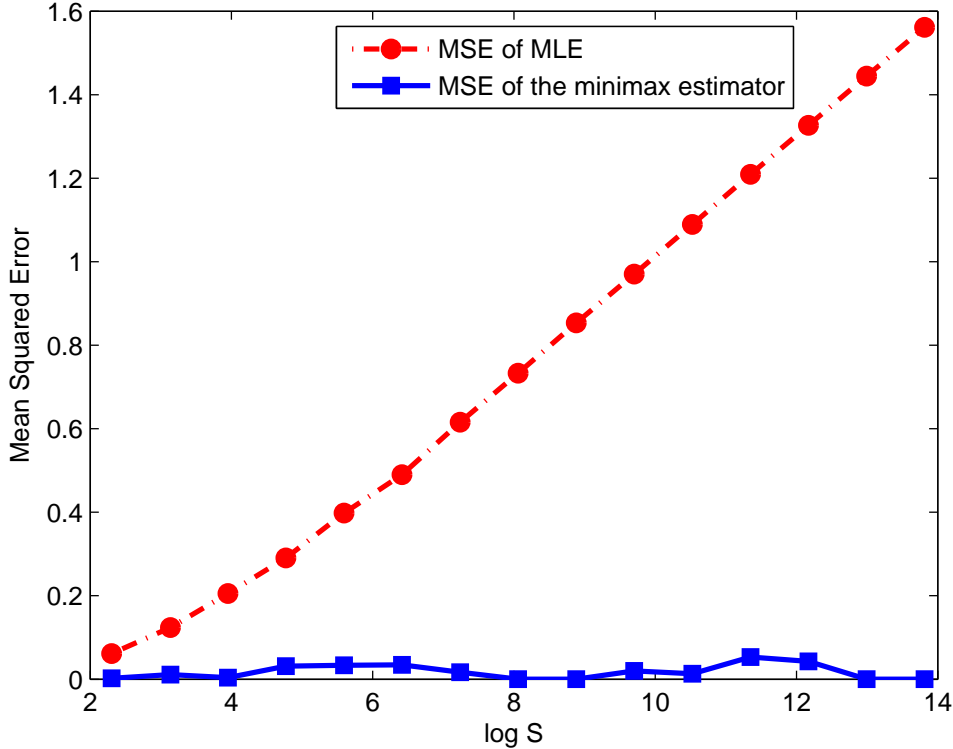}}
\caption{The empirical MSE of the estimator in \cite{Jiao--Venkat--Han--Weissman2014minimax} and the MLE along sequence $n = 5S/\ln S$, where $S$ is sampled equally spaced logarithmically from $10$ to $10^6$. The horizontal line is $\ln S$, and the vertical line is the MSE obtained using 20 Monte Carlo simulations from sampling a uniform distribution supported on $S$ elements. }\label{fig.MLE_ours_entropy}
\end{figure}

Specifically, for the estimation of the Shannon entropy, Valiant and Valiant~\cite{Valiant--Valiant2011} were the first to show that it is necessary and sufficient to take $n = \Theta(S/\ln S)$ samples. Later, in \cite{Valiant--Valiant2011power} they presented an improved estimator which achieves the optimal convergence rate. More recently, the present authors in \cite{Jiao--Venkat--Han--Weissman2014minimax}, and Wu and Yang in \cite{Wu--Yang2014minimax} independently developed schemes based on approximation theory that also achieve the optimal rates of convergence for the entropy.  In contrast to all these schemes which require $n = \Theta(S/\ln S)$ samples, the MLE requires $n = \Theta(S)$ samples~\cite{Paninski2003,Jiao--Venkat--Weissman2014MLE}. Figure~\ref{fig.MLE_ours_entropy} compares the performance of the essentially minimax optimal estimator in \cite{Jiao--Venkat--Han--Weissman2014minimax} and the MLE, and shows that this improvement can in fact be significant in practice. 

The substantial sub-optimality of the MLE is by no means particular to the entropy functional. \cite{Jiao--Venkat--Han--Weissman2014minimax}  demonstrated similarly substantial sub-optimality of the MLE in a wide family of functional estimation problems. In particular, an interesting observation therein is that the performance under $L_2$ risk of the optimal estimators with $n$ samples is essentially that of the MLE with $n \ln n$ samples. Intriguing as these findings are theoretically, they are valuable also to the practitioner encountering problems beyond functional estimation, as we illustrate next.

\subsection{Learning graphical models}

Given $n$ i.i.d. samples of a random vector $\mathbf{X} = (X_1,X_2,\ldots,X_d)$, where $X_i \in \mathcal{X}, |\mathcal{X}|<\infty$, we are interested in estimating the joint distribution of $\mathbf{X}$. It was shown that one needs to take $n = \Theta(|\mathcal{X}|^d)$ samples to consistently estimate the joint distribution \cite{Paninski2004variational}, which blows up easily as $d\to \infty$. Practically, it is convenient and necessary to impose some structure on the joint distribution $P_{\mathbf{X}}$ to reduce the required sample complexity. Chow and Liu \cite{Chow--Liu1968} considered this problem under the constraint that the joint distribution of $\mathbf{X}$ satisfies order-one dependence. To be precise, Chow and Liu assumed that $P_{\mathbf{X}}$ can be factorized as:
\begin{equation}
P_{\mathbf{X}} = \prod_{i = 1}^d P_{X_i|X_{\pi(i)}},
\end{equation}
where $X_{\pi(i)}$ denotes the parent node of $X_i$, and this dependence structure can be written as a tree with the random variables as nodes.

Towards estimating $P_{\mathbf{X}}$ from $n$ i.i.d. samples, Chow and Liu~\cite{Chow--Liu1968} considered solving for the MLE under the constraint that it factors as a tree. Interestingly, this optimization problem can be efficiently solved after being transformed into a Maximum Weight Spanning Tree (MWST) problem. Chow and Liu~\cite{Chow--Liu1968} showed that the MLE of the tree structure boils down to the following expression:
\begin{align}\label{eqn.CL}
E_{\mathrm{ML}} & = \arg\max_{E_Q: Q\textrm{ is a tree}} \sum_{e\in E_Q} I(\hat{P}_e),
\end{align}
where $I(\hat{P}_e)$ is the mutual information associated with the empirical distribution of the two nodes connected via edge $e$, and $E_Q$ is the set of edges of distribution $Q$ that factors as a tree. In words, it suffices to first compute the empirical mutual information between any two nodes (in total $\binom{d}{2}$ pairs), and the maximum weight spanning tree is the tree structure that maximizes the likelihood. To obtain estimates of distributions on each edge, Chow and Liu~\cite{Chow--Liu1968} simply assigned the empirical distribution. 

The Chow--Liu algorithm is widely used in machine learning and statistics as a tool for dimensionality reduction, classification, and as a foundation for algorithm design in more complex dependence structures~\cite{Zhou2011structure} in the theory of learning graphical models~\cite{Wainwright--Jordan2008,Koller--Friedman2009}. It has also been widely adopted in applied research, and is particularly popular in systems biology. For example, the Chow--Liu algorithm is extensively used in the reverse engineering of transcription regulatory networks from gene expression data~\cite{Meyer--Kontos--Lafitte--Bontempi2007information}. 

Considerable work has been dedicated to the theoretical properties of the CL algorithm. For example, Chow and Wagner~\cite{Chow--Wagner1973consistency} showed that the CL algorithm is consistent as $n\to \infty$. Tan et al.~\cite{Tan--Anandkumar--Tong--Willsky2011large} studies the large deviation properties of CL. However, no study justified the use of the CL in practical scenarios involving finitely many samples. As we elaborate in what follows, this is no coincidence, as the CL can be considerably improved on in practice. To explain the insights underlying our improved algorithm, we revisit equation~(\ref{eqn.CL}) and note that if we were to replace the empirical mutual information with the true mutual information, the output of the MWST would be the true edges of the tree. In light of this, the CL algorithm can be viewed as a ``plug-in'' estimator that replaces the true mutual information with an estimate of it, namely the empirical mutual information. Naturally then, it is to be expected that a better estimate of the mutual information would lead to smaller probability of error in identifying the tree. However, how bad can the empirical mutual information be as an estimate for the true mutual information? The following theorem implies that it can be highly sub-optimal in high dimensional regimes.

\begin{theorem}\label{thm.mutualsample}
Suppose we have two random variables $X_1,X_2\in \mathcal{X}, |\mathcal{X}|<\infty$. The minimax sample complexity in estimating the mutual information $I(X_1;X_2)$ under mean squared error is $\Theta(|\mathcal{X}|^2 / \ln |\mathcal{X}|)$, while the worse-case sample complexity required by the empirical mutual information to be consistent is $\Theta(|\mathcal{X}|^2)$.
\end{theorem}

The proof of Theorem~\ref{thm.mutualsample} is given in the appendix. It implies the essential optimality from a sample complexity viewpoint of the mutual information estimator proposed in~\cite{Jiao--Venkat--Han--Weissman2014minimax}. It is thus natural to suspect that using the latter in lieu of the empirical mutual information in the CL algorithm would lead to performance boosts. It is gratifying to find this intuition confirmed in all the experiments that we conducted. In the following experiment, we fix $d = 7,|\mathcal{X}| = 200$, construct a star tree (i.e. all random variables are conditionally independent given $X_1$), and generate a random joint distribution by assigning independent Beta$(1/2,1/2)$-distributed random variables to each entry of the marginal distribution $P_{X_1}$ and the transition probabilities $P_{X_k|X_1}, 2\leq k \leq d$ (with normalization). Then, we increase the sample size $n$ from $10^3$ to $2.6\times 10^4$, and for each $n$ we conduct $20$ Monte Carlo simulations. 

\begin{figure}[h]
\centerline{\includegraphics[width=0.8\linewidth]{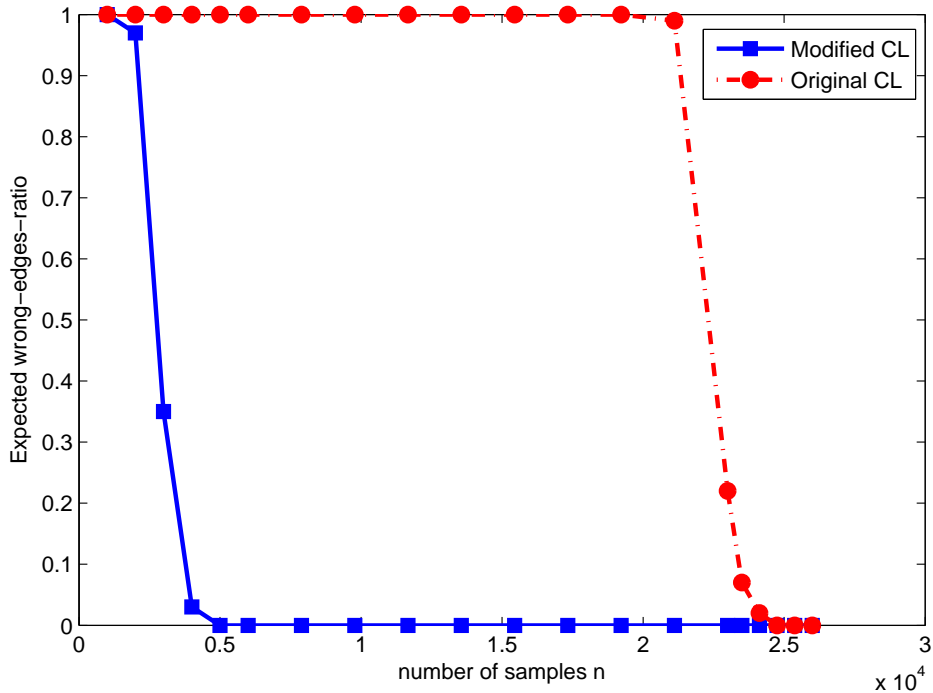}}
\caption{The expected wrong-edges-ratio of our modifed algorithm and the original CL algorithm for sample sizes ranging from $10^3$ to $2.6\times 10^4$.}\label{fig.CL_ours}
\end{figure}

Note that the true tree has $d-1 = 6$ edges, and any estimated set of edges will have at least one overlap with these $6$ edges because the true tree is a star graph. We define the wrong-edges-ratio in this case as the number of edges different from the true set of edges divided by $d-2 = 5$. Thus, if the wrong-edges-ratio equals one, it means that the estimated tree is maximally different from the true tree and, in the other extreme, a ratio of zero corresponds to perfect reconstruction. We compute the expected wrong-edges-ratio over $20$ Monte Carlo simulations for each $n$, and the results are exhibited in Figure~\ref{fig.CL_ours}.

Figure~\ref{fig.CL_ours} reveals intriguing phase transitions for both the modified and the original CL algorithm. When we have fewer than $1\times 10^3$ samples, both algorithms yield a wrong-edges-ratio of $1$, but soon after the sample size exceeds $5\times 10^3$, the modified CL algorithm begins to reconstruct the network perfectly, while the original CL algorithm continues to fail maximally until the sample size exceeds $25\times 10^3$, $5$ times the sample size required by the modified algorithm. The theoretical properties of these sharp phase transitions remain to be explored. 

\subsection{Bayesian network classifiers}
Given $n$ training samples, each of which has $d$ attributes $\mathbf{X}=(X_1,X_2,\cdots,X_d),X_i\in\mathcal{X}_i$ and a class label $C\in\mathcal{C}$, we are interested in constructing a classifier to assign a class label to a test instance characterized by its attributes. One important class of classifiers is the Bayes classifier \cite{Hastie--Tibshirani--Friedman2009elements}, which learns from training data the conditional joint distribution of $\bf X$ given the class label $C$. Then classification is done by applying the Bayes rule to compute the posterior probability of each class given the attribute vector. To estimate the conditional joint distribution, Friedman et al. \cite{Friedman--Geiger--Goldszmidt1997bayesian} assumed that $\bf X$ satisfies the order-one dependence conditioning on the class label, i.e., the joint probability of $\bf X$ given $C$ can be factorized into the product of the probabilities of each attribute conditioning on another attribute and the class label. To be precise, $P_{\mathbf{X}|C}$ can be factorized as
\begin{align}
  P_{\mathbf{X}|C} = \prod_{i=1}^d P_{X_i|X_{\pi(i)},C},
\end{align}
where, as in the preceding section, $X_{\pi(i)}$ denotes the parent node of $X_i$, and a tree graphical model can be established to describe this dependence structure.

In light of the CL algorithm, Friedman et al. \cite{Friedman--Geiger--Goldszmidt1997bayesian} proposed the tree-augmented naive Bayes (TAN) classifier. To construct the TAN classifier, the tree graphical model is established first using the CL algorithm, with a slight difference that the empirical mutual information $I(\hat{P}_e)$ in (\ref{eqn.CL}) is replaced by the conditional empirical mutual information $I(\hat{P}_e|C)$. Once the tree graphical model has been obtained, the empirical distributions $\hat{P}_C$ and $\hat{P}_{X_i|X_{\pi(i)},C}$ are used to estimate $P_C$ and $P_{\mathbf{X}|C}$, respectively, and both are substituted into 
\begin{align}
  f(\mathbf{x}) \triangleq \arg\max_{c\in\mathcal{C}} P_C(c)P_{\mathbf{X}|C}(\mathbf{x}|c),
\end{align}
which is the maximum a posteriori (MAP) estimator of the class label given attribute vector $\bf x$ using the Bayes rule.

Since we have demonstrated in the preceding section that the CL algorithm based on MLE is far from optimal, it is reasonable that we can harvest a performance gain in classification problems by simply using our better estimate of the mutual information for learning the tree graphical model. Specifically, we estimate the conditional mutual information via
\begin{align}
  I(\hat{P}_{X_iX_j}|C) = \hat{H}(X_i,C) + \hat{H}(X_j,C) - \hat{H}(C) - \hat{H}(X_i,X_j,C),
\end{align}
 where $\hat{H}$ is our improved entropy estimator. In this way, we construct a modified TAN classifier, and we remark that this construction does not impose an increased implementation burden since the computational complexity of the improved estimator $\hat{H}$~\cite{Jiao--Venkat--Han--Weissman2014minimax} is linear in the number of observations.

\begin{figure}[h]
\hspace*{0cm}
\centerline{\includegraphics[width=1.1\linewidth]{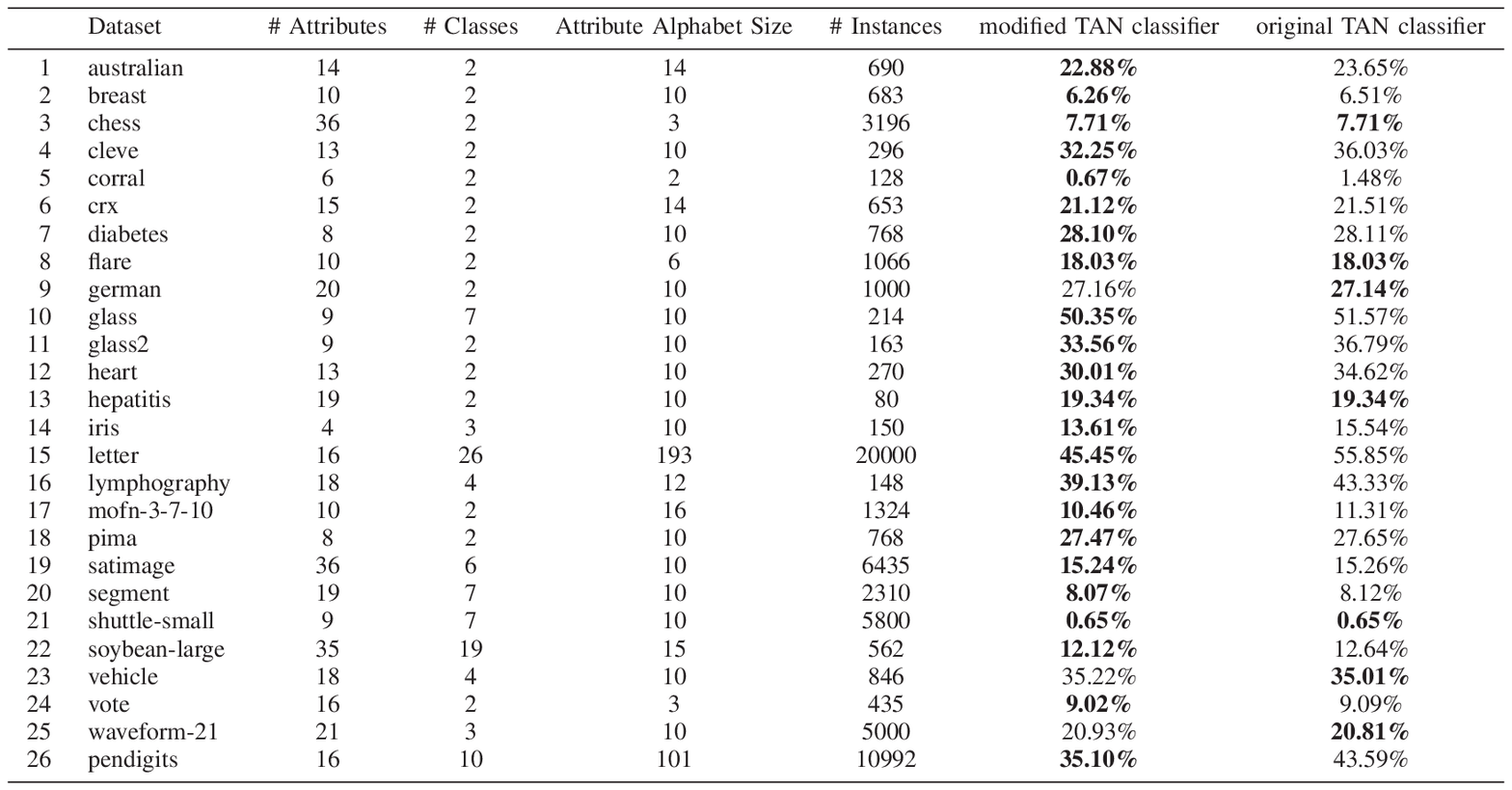}}
\caption{Description of the datasets used in experiments and along with classification errors in percentages. For all datasets with attributes taking continuous values, the 10-fold uniform quantization within its range is implemented. The attribute clustering is employed on datasets \emph{letter}, \emph{lymphography} and \emph{mofn-3-7-10}. In \emph{letter}, each clustered attribute contains 2 original attributes. In \emph{lymphography}, 9 clustered attributes contain respectively 2, 3, 3, 2, 2, 1, 1, 3, and 1 original attributes. In \emph{mofn-3-7-10}, 3 clustered attributes contain 3, 3, and 4 original attributes respectively. Minimum classification errors for each dataset are emphasized in bold. 
}
\label{fig.table1}
\end{figure}

Now we evaluate the performance gain of our modified classifiers in terms of the classification error via experimentation on a total of 26 datasets. All of the datasets are popular datasets from the UCI repository \cite{Murphy--Aha1995uci}, and are listed in Figure \ref{fig.table1} for reference. We note that the first 25 datasets are identical to those used in \cite{Friedman--Geiger--Goldszmidt1997bayesian}. Additionally, the  \emph{pendigits} dataset is selected since its attribute alphabet size, i.e. $S \triangleq \max_{1\le i\le d} |\mathcal{X}_i|$ is large. Recall that Theorem \ref{thm.mutualsample} indicates that for a large parameter dimension, the empirical mutual information can be highly sub-optimal for finite sample sizes. This is affirmed in the experimental results. 

\begin{figure}[h]
\hspace*{2cm}
\leftline{\includegraphics[width=0.8\linewidth]{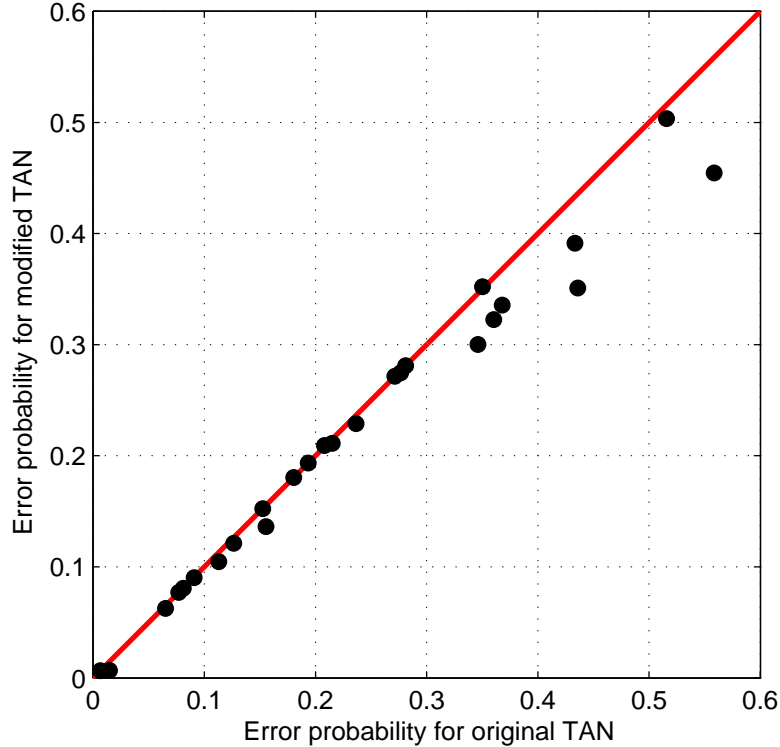}}
\caption{The scatter plot of the error probability comparing original TAN ($x$-axis) with the modified TAN classifier ($y$-axis). Points above the diagonal line corresponds to datasets where classical TAN classifier performs better, and points below the diagonal line corresponds to datasets where modified TAN classifier performs better.}\label{fig.TAN}
\end{figure}

First, we implement the 5-fold random cross validation repeatedly on all datasets for 100 times, and in each cross validation, the classification errors of original TAN and modified TAN classifiers are recorded separately. Figure~\ref{fig.table1} shows the mean values of all classification errors in percentage, where minimum classification errors for each dataset are in bold. For comparison, we investigate the classification error reduction using the modified classifiers via the scatter plot in Figure \ref{fig.TAN}.

%\begin{figure}[h]
%\leftline{\includegraphics[width=0.9\linewidth]{CL.eps}}
%\caption{The scatter plot comparing classical CL classifier ($x$-axis) with modified CL classifier ($y$-axis). In this scatter plot,
%points above the diagonal line corresponds to datasets where classical CL classifier performs better and points below the diagonal line corresponds to datasets
%where modified CL classifier performs better.}\label{fig.CL}
%\end{figure}

Figure \ref{fig.TAN} shows intriguing properties of the modified TAN classifier relative to the original one. Since none of the solid circles lies above the diagonal line, we conclude that our modified TAN classifier uniformly outperforms the original one in terms of classification errors. Furthermore, the top eight datasets with largest classification error reduction are \emph{letter}, \emph{pendigits}, \emph{heart}, \emph{lymphography}, \emph{cleve}, \emph{glass2}, \emph{iris} and \emph{glass}, which share a common feature that the squared maximum alphabet size is comparable to the number of observations, i.e., $S^2\cong n$. Also, in light of Theorem \ref{thm.mutualsample}, a remarkably lower risk in mutual information estimation is expected when the empirical mutual information begins to fail to be consistent. Hence, our experimental results are in accord with the theoretical findings that our improved estimator can estimate the mutual information consistently using reduced samples, which further results in lower classification errors.

\begin{figure}[h]
\hspace*{2cm}
\leftline{\includegraphics[width=0.8\linewidth]{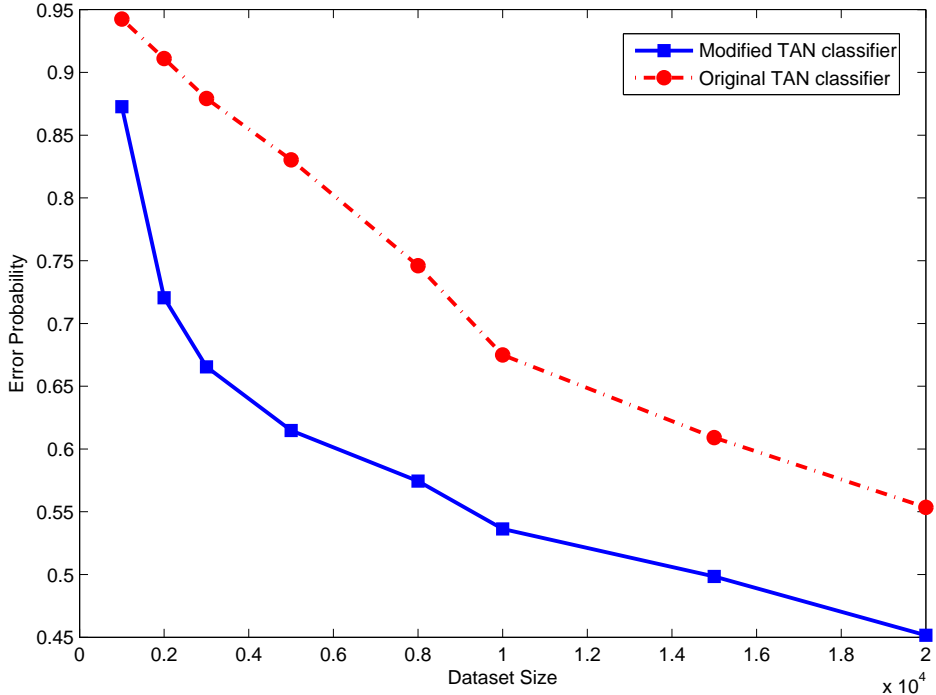}}
\caption{The error probability decay curve comparing original TAN classifier (dashed line) with modified TAN classifier (solid
line) in random subsets of the dataset letter. The $x$-coordinates of all squares (and circles) correspond to the subset size of 1000, 2000, 3000, 5000, 8000,
10000, 15000 and 20000, respectively.}\label{fig.letter}
\end{figure}

Second, to further convey the point that the modified classifiers require fewer samples to achieve an acceptable classification error, we conducted another experiment to compare the error probability decay curves under different classifiers. Specifically, sample sizes from 1000 to 20000 are selected, and for each sample size $n$, the preceding classification experiment on a training sample of size $4n/5$ and a testing sample of size $n/5$ is implemented 20 times, where in each time the training sample is a subset randomly generated from the training data in dataset \emph{letter}. Figure \ref{fig.letter} displays the relationship between the average classification errors and training sample sizes.

Figure \ref{fig.letter} exhibits a remarkable error reduction over the original scheme, uniformly over all sample sizes. For example, to achieve probability of error $0.7$, the sample size required by the modified TAN classifier is about 2000, while that for the original one is about 10000. Note that since there are 26 class labels in the dataset \emph{letter} and the fraction for each class does not exceed 5\%, a random guess would result in at least classification error 95\%.

We remark that some classification errors listed in Figure \ref{fig.table1} are worse than those in \cite{Friedman--Geiger--Goldszmidt1997bayesian}, as we did not adopt the smoothing method employed in~\cite{Friedman--Geiger--Goldszmidt1997bayesian}. That method reduces the dependence between the tree graphical model and classification results significantly, and additional experiments (not reported here) show that the modified classifier still uniformly outperforms the original one when the smoothing method are applied in both classifiers.

\section{Concluding Remarks}

We demonstrated both theoretically and empirically, that automatic use of MLE without justification may result in highly sub-optimal performance.  Therefore, we alert applied scientists to the fact that even if the MLE seems to have taken into consideration all the \emph{prior knowledge} about the problem (such as the CL algorithm), it still might be significantly improved upon. This effect is far wider than we had the space to demonstrate here, and cognizance of it may lead to significant performance boosts in various disciplines. For example, our methodology may provide better estimators for the directed information~\cite{Massey90, Jiao--Permuter--Zhao--Kim--Weissman2013}. 

\section{Acknowledgments} 
This work was partially supported by two Stanford Graduate Fellowships and NSF grant CCF-0939370. We thank Dmitri Serguei Pavlichin for interesting discussions related to the literature on applications of maximum likelihood in biology and physics. We thank Ritesh Kolte for suggesting the use of the wrong-edges-ratio metric in our evaluation of the modified CL algorithm over the original one. 

%----------------------------------------------------------------------------------------
%	APPENDICES (OPTIONAL)
%----------------------------------------------------------------------------------------

\appendix[Proof of Theorem~\ref{thm.mutualsample}]
The mutual information between two random variables $X_1,X_2$ can be written as
\begin{equation}\label{eqn.expand}
I(X_1;X_2) = H(X_1) + H(X_2) - H(X_1,X_2),
\end{equation}
where $H(\cdot)$ is the entropy functional defined as $H(P) = \sum_{i = 1}^{|\mathcal{X}|}- p_i \ln p_i$. It was shown in \cite{Valiant--Valiant2011, Valiant--Valiant2011power, Jiao--Venkat--Han--Weissman2014minimax, Wu--Yang2014minimax} that the sample complexity for estimating $H(P)$ is $\Theta(|\mathcal{X}|/\ln |\mathcal{X}|)$. Applying this result in estimating $I(X_1;X_2)$, we know it suffices to take $\Omega(|\mathcal{X}|^2/\ln |\mathcal{X}|)$ samples to consistently estimate $I(X_1;X_2)$. It is also the optimal sample complexity as we argue next. Suppose one can construct an estimator $\hat{I}$ that can consistently estimate $I(X_1;X_2)$ using $o(|\mathcal{X}|^2/\ln |\mathcal{X}|)$ samples. Then one can construct an estimator for $H(X_1,X_2)$ using $\hat{H}(X_1) + \hat{H}(X_1) - \hat{I}$,  with $\hat{H}$ denoting the optimal entropy estimator, which is a consistent estimator for $H(X_1,X_2)$ with $o(|\mathcal{X}|^2/\ln |\mathcal{X}|)$ samples. It then violates the lower bound for estimating entropy.

At the same time, it was shown in \cite{Paninski2003, Jiao--Venkat--Weissman2014MLE} that it is necessary and sufficient for the MLE to take $\Theta(|\mathcal{X}|)$ samples to consistently estimate $H(X)$. Obviously, (\ref{eqn.expand}) shows that $\Omega(|\mathcal{X}|^2)$ samples suffice for the empirical mutual information to be consistent. Now we argue that it is also necessary. Suppose that the empirical mutual information is consistent with $o(|\mathcal{X}|^2)$ samples. Then (\ref{eqn.expand}) implies that the empirical entropy estimator for $H(X_1,X_2)$ is also consistent with $o(|\mathcal{X}|^2)$ samples, which violates the results in \cite{Paninski2003, Jiao--Venkat--Weissman2014MLE}.

%----------------------------------------------------------------------------------------
%	ACKNOWLEDGEMENTS
%----------------------------------------------------------------------------------------

%----------------------------------------------------------------------------------------
%	BIBLIOGRAPHY
%----------------------------------------------------------------------------------------

\bibliographystyle{IEEEtran}
\bibliography{di}

\end{document}